\documentstyle[11pt,twoside,epsfig]{article}
\newif\ifamsfonts
\amsfontstrue \ifamsfonts
\font\twlbbb=msbm10 scaled\magstep1 \font\egtbbb=msbm8
\font\sixbbb=msbm6
\newfam\bbbfam
\textfont\bbbfam=\twlbbb \scriptfont\bbbfam=\egtbbb
\scriptscriptfont\bbbfam=\sixbbb

\else
\fi

\newtheorem{thm}{Theorem}
\makeatletter\@addtoreset{equation}{section}\makeatother

\newcommand{\sech}{\mathop{\rm sech}\nolimits}

\newcommand{\abs}[1]{\left | #1 \right |}

\newcommand{\id}{\,{\rm d}}

\newcommand{\iu}{\mskip2mu{\rm i}\mskip1mu}

\def\pt{\partial}
\def\bd{\begin{displaymath}}
\def\ed{\end{displaymath}}
\def\bqns{\begin{eqnarray*}}
\def\bi{\begin{itemize}}
\def\ei{\end{itemize}}
\def\beq{\begin{quote}}
\def\eeq{\end{quote}}
\def\ben{\begin{enumerate}}
\def\een{\end{enumerate}}
\def\eqns{\end{eqnarray*}}
\def\bq{\begin{equation}}
\def\bqn{\begin{eqnarray}}
\def\eq{\end{equation}}
\def\eqn{\end{eqnarray}}

\def\e{\varepsilon}
\def\l{\lambda}

\def\D{\Delta}
\def\o{\omega}

\def\di{\displaystyle}

\newcounter{saveeqn}

\begin{document}
\title{Stability of Waves in Multi-component DNLS system}
%\footnote{}

\author{V. M. Rothos$^\dagger$\footnote{On leave: Department of Applied Mathematics,
University of Crete, Heraklio, GR71409 Greece} and P.G. Kevrekidis$^{\dagger \dagger}$,
\\
{\small $^{\dagger}$ Department of Mathematics, Physics
Computational Sciences,} \\ {\small Faculty of Technology,
Aristotle University of Thessaloniki,
 Thessaloniki 54124 Greece} \\
{\small $^{\dagger \dagger}$ Department of Mathematics and Statistics,
University of Massachusetts
Amherst, MA 01003-4515
USA} }

\maketitle
\begin{abstract}
In this work, we systematically generalize the Evans function methodology
to address vector systems of discrete equations. We physically motivate
and mathematically use as our case example a vector form of the discrete
nonlinear Schr{\"o}dinger equation with both nonlinear and linear couplings
between the components. The Evans function allows us to qualitatively
predict the stability of the nonlinear waves under the relevant perturbations
and to quantitatively examine the dependence of the corresponding point
spectrum eigenvalues on the system parameters. These analytical predictions
are subsequently corroborated by numerical computations.
\end{abstract}
\begin{itemize}
\item[] 2000 MSC 34C, 37K, 39 \item[] PACS 0230K, 0240V, 0545
\end{itemize}
\newpage
\tableofcontents\newpage
%%%%%%%%%%%%%%%%%%%%%%%%%%%%%%%%%%%%%%%%%%%%%%%%%%%%%%%%%%%%%%%%%%%%%%
%
%                        Section 1
%%%%%%%%%%%%%%%%%%%%%%%%%%%%%%%%%%%%%%%%%%%%%%%%%%%%%%%%%%%%%%%%%%%%%%

\section{Introduction}
\setcounter{equation}{0}

In the last few years, the study of solitary waves in
multi-component systems has drawn a large focus of attention both
in the physics of Bose-Einstein  condensates
\cite{books,books1,books2,reviews,reviews1,reviews2,reviews3}, and
in that of nonlinear optical fiber and waveguide arrays
\cite{agrawal,variational}.

In the case of BEC dynamics, if the condensates are in a deep optical
lattice, then their dynamics can be described by a discrete nonlinear
Schr{\"o}dinger (DNLS) equation \cite{smerzi,konotop,alfimov}. Multi-species
condensates can arise, in this setting, as mixtures of different spin states in
$^{87}$Rb \cite{myatt,dsh} and $^{23}$Na \cite{stamper}
condensates. Furthermore, theoretical studies have discussed
the possibilities of mixtures between different atomic species,
such as Na--Rb
\cite{pu1,pu2}, K--Rb \cite{simoni1,simoni2}, Cs--Rb \cite{jam}
and Li--Rb \cite{LiRb}. A two-species BEC, in fact in a  $^{41}$K--$^{87}$Rb
mixture, has been recently reported
\cite{KRb}; finally, a mixture of $^{7}$Li and $^{133}$Cs was also
experimentally investigated \cite{LiCs} (albeit in a non-condensed state).
In the above contexts, the relevant theoretical models, in the presence
of an optical lattice trapping \cite{reviews}, consists of one
DNLS equation per atomic species. These equations are  coupled by nonlinear
cross-phase modulation (XPM); however, linear coupling terms can also be
applied to the description of BEC dynamics in this mean-field approximation.
The nonlinear interaction between the
components is generated by (inter-species)
atomic collisions, while linear coupling may be
readily induced by an external microwave or radio-frequency field which
induces Rabi \cite{NewZealand} or Josephson \cite{BEC-Josephson}
oscillations between populations of the two states.

Numerous issues have been
considered, especially in the continuum (i.e., non-lattice)
counterpart of the above framework. Among them, are ground-state
solutions \cite{pu1,shenoy,esry}, small-amplitude excitations
\cite{pu2,excit}, formation of \textit{domain walls} (DWs) between
immiscible species \cite{Marek,healt,prop,Mehrasin}, bound states of
dark-bright \cite{anglin} and dark-dark \cite{obsantos},
dark-gray, bright-gray, bright-antidark and dark-antidark
\cite{epjd} complexes of solitary waves, spatially periodic states
\cite{decon} and modulated amplitude waves \cite{mason} among others.
%including an effect on DWs in the case of immiscibility
%\cite{Merhasin}.

A realization of the above discussed system is quite possible also
in the setting of the coupled optical waveguides.
In that case, the two species represent either two orthogonal
polarizations of light, or two signals with different carrier wavelengths.
In the latter case, the linear coupling can also be
implemented, by twisting the waveguides or elliptically deforming them, in
the case of linear or circular polarizations, respectively.
In this context, strongly localized vector
(two-component) discrete solitons have been identified,
%in models of
%nonlinear waveguide arrays, where two fields interact
%through XPM,
both in one-dimension \cite{b16,b17}, as well as in two dimensions
\cite{jared}. Similarly to their continuous counterparts \cite
{b18}, %b19,b20,b21,b22}
 these vector solitons may have components
of different types (bright, dark, or antidark). In particular,
symbiotic bright-dark and dark-antidark pairs were predicted in
such systems \cite {b16,b17}. Such solitary waves
have also been observed experimentally, see e.g., \cite{meier}.

These recent developments, on both theoretical and also
experimental aspects of such multi-component systems underscore the
relevance of further systematic studies of their coherent structures
and, importantly, also of their stability. More specifically, in the
discrete setting, the stability of
solitary waves in (a single-component)
DNLS and related models was addressed in \cite{kev5}.
However, to the best of our knowledge, there were no works developing
techniques systematically
addressing the stability of waves in the context of vector discrete equations.
This formulates the problem that the present study aims at addressing.
In particular, we use as our starting point a vector integrable
discretization of \cite{ablowitz,trub1} (see also \cite{wadati} and
references therein). We then include perturbations, through the inclusion of
physically relevant terms, such as ones that appear in the
framework of the coupled DNLS models or the linear coupling
discussed above (see e.g. \cite{NewZealand,BEC-Josephson}). We focus
on the influence of such terms on the point spectrum
eigenvalues of the linearization around the perturbed solitary waves.
To monitor the latter, we generalize the Evans function methodology
of \cite{kev5} (see also \cite{kapitula99} for the continuum setting)
to address
the vector case and obtain good agreement of the relevant predictions
with the qualitative phenomenology and the quantitative dependence
on (the perturbed) system parameters.

Our presentation will be structured as follows. In section 2, we are
going to give the general setup and notation of the coupled equation
system. In section 3, we are going to develop the Evans function for
the vector case and use it to obtain the eigenvalues in the case of
DNLS-like perturbations. In section 4, we are going to give the corresponding
results in the presence of linear coupling. In these two sections, the
analytical results will be complemented with numerical computations.
Finally, in section 5, we summarize our findings and present our
conclusions.

\section{The CDNLS equation and Notation}
We consider the case of two coupled discrete NLS (CDNLS) equations in the form
\begin{eqnarray}
{\iu}{\dot
  u}_{n}&=&-{\D}_{2}u_{n}-(\abs{u_{n}}^{2}+\abs{v_{n}}^{2})(u_{n+1}+u_{n-1}) - 2 {\e}
u_{n}(\abs{u_{n}}^{2}+\abs{v_{n}}^{2})\cr\cr {\iu}{\dot
  v}_{n}&=&-{\D}_{2}v_{n}-(\abs{v_{n}}^{2}+\abs{u_{n}}^{2})(v_{n+1}+v_{n-1}) - 2 {\e}
v_{n}(\abs{v_{n}}^{2}+\abs{u_{n}}^{2}) \label{cdnls}
\end{eqnarray}
with $${\D}_{2}:=\frac{1}{h^{2}}(u_{n+1}-2u_{n}+u_{n-1})$$; $h$
is the step-size of discretization and $n$ is the lattice site index.
Our motivating physical setting stems from the consideration of 
coupled hyperfine states of an atomic species (such as the spin states
$|1,-1>$ and $|2,1>$ of $^{87}Rb$; see \cite{myatt,dsh} 
for relevant details), examined in the presence of a
quasi-one-dimensional deep optical lattice potential. A vector
generalization of the derivation of \cite{alfimov} would establish
the coupled discrete NLS model 
(with near-unity nonlinearity coefficients \cite{dsh})
as the appropriate underlying mathematical framework. However, a
starting point that is more tractable analytically is the 
$\e=0$ variant of the above model of Eqs. (\ref{cdnls})
 (i.e., the ``Ablowitz-Ladik limit'' \cite{ablowitz,trub1}).   
We therefore study first the existence
and stability of discrete solitons for CDNLS equations starting
from the $\e=0$ limit and using
continuation in $\e$ small to examine the persistence of such
solutions in the non-integrable limit with the onsite nonlinearity. 
We study analytically the
stability of discrete solitons of the above coupled system for
$C \equiv 1/h^2=1$ through the Evans method and the properties of reduced
discrete dynamical system (perturbed 4-dimensional mapping). Based
on the integrability of this map one can provide an analytic
expression of the discrete Evans function. Concerning small
perturbations of integrable CDNLS model, we wish to study the
stability criteria for discrete solitons and test these
predictions against numerical simulations.
We note in passing that this type of discretization does have the
relevant Manakov continuum limit as $h \rightarrow 0$, where the equations
become 
\begin{eqnarray}
i u_t =- \Delta u - (1+\epsilon) \left( |u|^2 + |v|^2 \right) u,
\label{cont1}
\\
i v_t =- \Delta v - (1+\epsilon) \left( |u|^2 + |v|^2 \right) v,
\label{cont2}
\end{eqnarray} 
up to rescaling of the amplitudes ($u \rightarrow u \sqrt{1+ \epsilon}$
and $v \rightarrow v \sqrt{1+ \epsilon}$) or one of space and time
(i.e., $x \rightarrow x \sqrt{1+\epsilon}$
and $t \rightarrow t (1+\epsilon)$). Notice that this particular
discretization is motivated by the wide volume of work on the
so-called Salerno model \cite{salerno} which is also considering
a mixed nonlinearity of combined local and nearest-neighbor 
nonlinear term.

%Finally, the relevance
%of the existence and stability of such solutions to presently
%existing experimental settings will be briefly commented upon.
%both in the nonlinear optics
%context, as well as in the one of atomic physics will be highlighted
%by giving physical estimates of parameters for which such solitary
%waves can be observed.

We are interested in stationary solutions of the system of
equations (\ref{cdnls}) and make the ansatz

\bq
u_{n}(t)=q_{n}{\rm e}^{-{\iu}{\o}t}, \qquad v_{n}(t)=p_{n}{\rm
e}^{-{\iu}{\o}t}, \label{stationary} \eq

with a uniform rotation frequency $\o$. Using Equations
(\ref{stationary}) in Equations (\ref{cdnls}), we
arrive at two coupled second--order difference equations

\bqn
q_{n+1}+q_{n-1}&=&\frac{\di{2-{\o}{h^{2}}}}{\di{1+{h^{2}}(q^{2}_{n}+p^{2}_{n})}}\,q_{n}
-{\e}{2h^{2}}q_{n}\frac{\di{(q^{2}_{n}+p^{2}_{n})}}{\di{1+{h^{2}}(q^{2}_{n}+p^{2}_{n})
}}\cr\cr
p_{n+1}+p_{n-1}&=&\frac{\di{2-{\o}{h^{2}}}}{\di{1+{h^{2}}(q^{2}_{n}+p^{2}_{n})}}\,p_{n}
-{\e}{2h^{2}}p_{n}\frac{\di{(q^{2}_{n}+p^{2}_{n})}}{\di{1+{h^{2}}(q^{2}_{n}+p^{2}_{n})
}}. \label{map1} \eqn

For $\e=0$, we obtain the integrable standard--like map with two
invariants:
\[
I_{1}=h^{2}\,\bigg[\,\Big(\,q^{2}_{n}+q^{2}_{n-1}-2{\rm
c}q_{n}q_{n-1}\,\Big)+\Big(\,p^{2}_{n}+p^{2}_{n-1}-2{\rm
c}p_{n}p_{n-1}\,\Big)+h^{2}\,(q^{2}_{n}+p^{2}_{n})(q^{2}_{n-1}+p^{2}_{n-1})\,\bigg]
\]
\[
I_{2}=\frac{h^{2}}{{\rm
c}^{2}}\,\bigg[\,q^{2}_{n}+q^{2}_{n-1}-2{\rm
c}q_{n}q_{n-1}+p^{2}_{n}+p^{2}_{n-1}-2{\rm
c}p_{n}p_{n-1}+\frac{h^{2}}{{\rm c}}\,\Big(\,
q_{n}q_{n-1}+p_{n}p_{n-1}\,\Big)^{2}\,\bigg]
\]
where ${\rm c}=2-{\o}h^{2}$.

Upon setting
\[
r_{n}=(q_{n}-q_{n-1})/h,\quad s_{n}=(p_{n}-p_{n-1})/h,
\]
the steady--state problem for CDNLS can be formulated as

\bqn q_{n+1}&=&({\rm
c}{\chi}-1)q_{n}+r_{n}h+{\e}\big(\,-{2h^{2}}\,\big)q_{n}(q^{2}_{n}+p^{2}_{n}){\chi}\cr\cr
r_{n+1}&=&({\rm
c}{\chi}-1)\frac{q_{n}}{h}+r_{n}+{\e}\big(\,-{2h}\,\big)q_{n}(q^{2}_{n}+p^{2}_{n}){\chi}\cr\cr
p_{n+1}&=&({\rm
c}{\chi}-1)p_{n}+s_{n}h+{\e}\big(\,-{2h^{2}}\,\big)p_{n}(q^{2}_{n}+p^{2}_{n}){\chi}\cr\cr
s_{n+1}&=&({\rm
c}{\chi}-1)\frac{p_{n}}{h}+s_{n}+{\e}\big(\,-{2h}\,\big)p_{n}(q^{2}_{n}+p^{2}_{n}){\chi},
\label{map2} \eqn where
\[
{\chi}=\frac{1}{1+h^{2}(q^{2}_{n}+p^{2}_{n})}.
\]
The equations (\ref{map2}) are written so that when $\e=0$, they are
then exactly the steady--state problem for integrable CDNLS.

When $\e=0$ the solutions are given by \bqn
Q_{n}(\xi)&=&\frac{{\alpha}_{1}{\sinh}2W
}{\sqrt{{\alpha}^{2}_{1}+{\alpha}^{2}_{2}}}\sech({\tilde
W}n+\xi),\cr
P_{n}(\xi)&=&\frac{{\alpha}_{2}{\sinh}2W}{\sqrt{{\alpha}^{2}_{1}+{\alpha}^{2}_{2}}}\sech({\tilde
W}n+\xi), \label{solutions} \eqn with ${\tilde
W}={\cosh}^{-1}({\rm c}/2)=2 W$. For small $h$, $\tilde{W} \approx
\sqrt{-\omega} h$. Note that these solutions precisely describe
the stable and unstable manifolds of the fixed point $(0,0,0,0)$,
and that these manifolds intersect non--transversely. This is a
non--generic phenomenon for standard-like maps, and hence it is
expected that the intersection, if it persists, will be transverse
for $\e>0$.

Addressing the linear stability of the wave, we set
$u_{n}=u_{_{R}}+{\iu}u_{_{I}}, v_{n}=v_{_{R}}+{\iu}v_{_{I}}$ and
linearizing CDNLS about the wave, one has the linearized problem
\bqn {\pt}_{t}u_{_{R}}&=& -L_{_{-}}u_{_{I}}\cr
{\pt}_{t}u_{_{I}}&=& L_{_{Q+}}u_{_{R}}+L_{_{Q}}v_{_{R}}\cr
{\pt}_{t}v_{_{R}}&=& -L_{_{-}}v_{_{I}}\cr {\pt}_{t}v_{_{I}}&=&
L_{_{P}}u_{_{R}}+L_{_{P+}}v_{_{R}}
 \label{lin-cdnls} \eqn

Hence, the operators $L_{_{-}}, L_{_{Q}}, L_{_{P}}, L_{_{Q+}},
L_{_{P+}}$ are given by \bqn \label{operators}
L_{_{-}}&=&{\Delta}_{2}+{\omega}+(Q^{2}_{n}+P^{2}_{n})({\rm
e}^{\pt}+{\rm e}^{-\pt})\cr\cr
L_{_{Q+}}&=&{\Delta}_{2}+{\omega}+2(Q_{n+1}+Q_{n-1})Q_{n}+(Q^{2}_{n}+P^{2}_{n})({\rm
e}^{\pt}+{\rm e}^{-\pt})\cr\cr
L_{_{Q}}&=&2P_{n}(Q_{n+1}+Q_{n-1})\cr\cr
L_{_{P+}}&=&{\Delta}_{2}+{\omega}+2(P_{n+1}+P_{n-1})P_{n}+(Q^{2}_{n}+P^{2}_{n})({\rm
e}^{\pt}+{\rm e}^{-\pt})\cr\cr L_{_{P}}&=&2Q_{n}(P_{n+1}+P_{n-1})
\eqn where ${\rm e}^{\pm\pt}u_{n}=u_{n\pm 1}$. Upon setting \[
\textbf{L}=\left(%
\begin{array}{cccc}
  0& -L_{_{-}} & 0 & 0 \\
  L_{_{Q+}} & 0 & L_{_{Q}} & 0 \\
  0 & 0 & 0 & -L_{_{-}} \\
  L_{_{P}} & 0 & L_{_{P+}} & 0 \\
\end{array}%
\right)
\]
it is not difficult to check that
\[
\textbf{L}\left(%
\begin{array}{c}
  {\pt}_{\omega}Q_{n} \\ 0\\
  {\pt}_{\omega} P_{n} \\
  0 \\
\end{array}%
\right)=\left(%
\begin{array}{c}
  0 \\
  Q_{n} \\
  0\\
  P_{n} \\
\end{array}
\right),\qquad \textbf{L}\left(%
\begin{array}{c}
  0 \\
  Q^{c}_{n} \\ 0\\
   P^{c}_{n}\\
\end{array}%
\right)=\left(%
\begin{array}{c}
  {\pt}_{\xi} Q_{n}  \\ 0 \\
  {\pt}_{\xi} P_{n} \\
  0 \\
\end{array}
\right).
\]
$Q^{c}_{n}, P^{c}_{n}$ satisfy the condition
\[
\lim_{h\to 0^{+}}\left(%
\begin{array}{c}
   Q^{c}_{n}\\
  P^{c}_{n}\\
\end{array}
\right)=-\frac{1}{2}x\left(%
\begin{array}{c}
   Q(x)\\
  P(x)\\
\end{array}
\right),
\]
where $Q(x), P(x)$ are the continuum limit of $Q_{n}(\xi),
P_{n}(\xi)$. The eigenvalue at $\lambda=0$ has
%geometric
%multiplicity four and
algebraic multiplicity six.

\section{Evans Function and stability}

We rewrite the eigenvalue problem $(\textbf{L}-\l)\textbf{u}=0$ as
a system of difference equations \bq Y_{n+1}=\textbf{A}(\l, n)Y_{n}
\label{lin1} \eq It is known that there exist solution sets
${\textbf{Y}^{1}_{n}, \textbf{Y}^{2}_{n},\textbf{Y}^{3}_{n}}$ and
${\textbf{Y}^{4}_{n}, \textbf{Y}^{5}_{n},\textbf{Y}^{6}_{n}}$ to
equation (\ref{lin1}) such that $\abs{\textbf{Y}^{i}_{n}}\to 0$
exponentially fast as $n\to -\infty$ for $i=1, 2, 3$ and
$\abs{\textbf{Y}^{i}_{n}}\to 0$ exponentially fast as $n\to
\infty$ for $i=4, 5, 6$.

The Evans function associated with (\ref{lin1}) is given by
\[
E(\l)=(\textbf{Y}^{-}_{n}\wedge
\textbf{Y}^{+}_{n})/\prod_{j=0}^{n-1}\det(\textbf{A}(\l, j))
\]
satisfies the following properties \cite{kev5}: $E(\l)$ is
analytic in $\l$, $E(\l)=0$ iff equation (\ref{lin1}) has a
bounded solution and the order of the zero is equal to the
algebraic multiplicity of the eigenvalue.

The adjoint problem associated with equation (\ref{lin1}) is \bq
\textbf{Z}_{n+1}=[\textbf{A}(\l, n)^{-1}]^{*}\textbf{Z}_{n}
\label{adjlin1} \eq and its solution satisfies:
\[
\textbf{Y}^{i}_{n}\cdot\textbf{Z}^{j}_{n}=\delta_{ij} \quad
i,j=1,2,3,4,5,6.
\]
Assuming that $E(\l)\neq 0$, these solutions furthermore have the
property $\abs{\textbf{Z}^{i}_{n}}\to 0$ exponentially fast as
$n\to \infty$ for $i=1, 2, 3$ and $\abs{\textbf{Z}^{i}_{n}}\to 0$
exponentially fast as $n\to -\infty$ for $i=4, 5, 6$.

For the particular problem of CDNLS, we obtain the linearized
system (\ref{lin1}) with \bq \label{lin1CDNLS} \textbf{A}(\l, n)=\left(%
\begin{array}{cccccccc}
  \gamma_{_{Q}} & \l \alpha & h & 0 & -L_{_{P}}\alpha & 0 & 0 & 0 \\
   -\lambda\alpha & (\beta\alpha-1)& 0 & h & 0 & 0 & 0 & 0\\
  {\gamma_{_{Q}}-1}/{h} & {\l\alpha}/{h} & 1 & 0 &  -L_{_{P}}{\alpha}/{h}& 0 & 0 &
  0  \\
  {-\lambda\alpha}/{h} & {(\beta\alpha-2)}/{h} & 0 & 1& 0 & 0 & 0 & 0 \\
  -L_{_{Q}}\alpha  & 0 & 0 & 0 & \gamma_{_{P}} & \lambda\alpha & h & 0 \\
  0 & 0 & 0 & 0 & -\lambda\alpha & (\beta\alpha-1) & 0 & h \\
  -L_{_{Q}}{\alpha}/{h} & 0 & 0 & 0 & {\gamma_{_{P}}-1}/{h} & {\lambda\alpha}/{h} & 1 & 0 \\
  0 & 0 & 0 & 0 & {-\lambda\alpha}/{h} & (\beta\alpha-2)& 0 & 1 \\
\end{array}%
\right) \eq where
\[
\alpha=\Big(\frac{1}{h^{2}}+(P^{2}_{n}+Q^{2}_{n})\Big)^{-1},\quad
\beta=\frac{2}{h^{2}}-\omega,
\]
\[
\gamma_{_{Q}}=\alpha(\beta-2Q_{n}(Q_{n+1}+Q_{n-1})-1), \quad
\gamma_{_{P}}=\alpha(\beta-2P_{n}(P_{n+1}+P_{n-1})-1).
\]

The solutions of (\ref{lin1}) about $\l=0$  are the following \bqn
\label{lin1-solutions}
\textbf{Y}^{1,-}_{n}(0)=\textbf{Y}^{1,+}_{n}(0)&=&\Big[\,\pt_{\xi}Q_{n},
0, \pt_{\xi}(Q_{n}-Q_{n-1})/h, 0, \pt_{\xi}P_{n}, 0,
\pt_{\xi}(P_{n}-P_{n-1})/h, 0\,\Big]^{\top}\cr\cr
\textbf{Y}^{2,-}_{n}(0)=\textbf{Y}^{2,+}_{n}(0)&=&\Big[\,0, Q_{n},
0, (Q_{n}-Q_{n-1})/h, 0, 0, 0, 0\,\Big]^{\top}\cr\cr
\textbf{Y}^{3,-}_{n}(0)=\textbf{Y}^{3,+}_{n}(0)&=&\Big[\,0, 0, 0,
0,P_{n}, 0, (P_{n}-P_{n-1})/h, 0\,\Big]^{\top} \eqn

Furthermore, define the relevant adjoint solutions as \bqn
\label{adjlin1-solutions}
\textbf{Z}^{1}_{n}&=&\Big[\,\pt_{\xi}(Q_{n-1}-Q_{n})/h, 0,
\pt_{\xi}Q_{n}, 0, \pt_{\xi}(P_{n-1}-P_{n})/h, 0, \pt_{\xi}P_{n},
0 \,\Big]^{\top}\cr\cr \textbf{Z}^{2}_{n}&=&\Big[\, 0,
(Q_{n-1}-Q_{n})/h, 0, Q_{n}, 0, 0, 0, 0, 0 \,\Big]^{\top}\cr\cr
\textbf{Z}^{3}_{n}&=&\Big[\, 0, 0, 0, 0, (P_{n-1}-P_{n})/h, 0,
P_{n}, 0 \,\Big]^{\top}. \eqn

Finally for the problem $\textbf{Y}_{n}=\textbf{A}(0,
n)\textbf{Y}$ define the solutions:
\[
u^{1}_{n}=\textbf{Y}^{1,-}_{n}(0),\quad
u^{2}_{n}=\textbf{Y}^{2,-}_{n}(0),\quad
u^{3}_{n}=\textbf{Y}^{3,-}_{n}(0)
\]
and let $u^{4}_{n}, u^{5}_{n}$ and $u^{6}_{n}$ satisfy  the
relations
\[
u^{4}_{n}\cdot \textbf{Z}^{1}_{n}=0,\quad u^{5}_{n}\cdot
\textbf{Z}^{1}_{n}=1,\quad u^{6}_{n}\cdot \textbf{Z}^{1}_{n}=0
\]
\[
u^{4}_{n}\cdot \textbf{Z}^{2}_{n}=1,\quad u^{5}_{n}\cdot
\textbf{Z}^{2}_{n}=0,\quad u^{6}_{n}\cdot \textbf{Z}^{2}_{n}=1
\]
\[
u^{4}_{n}\cdot \textbf{Z}^{3}_{n}=1,\quad u^{5}_{n}\cdot
\textbf{Z}^{3}_{n}=0,\quad u^{6}_{n}\cdot \textbf{Z}^{3}_{n}=0.
\]

The equation of variation with respect to $\e$ for the stable
$W^{s}$ and $W^{u}$ manifolds is given by the non-homogeneous
problem \bq \textbf{Y}_{n+1}= \textbf{A}(\l,
n)\textbf{Y}_{n}+\textbf{g}_{n} \label{nonhom} \eq where
$\{\,\textbf{g}_{n}\,\}$ is a uniformly bounded sequence:
\[
\textbf{g}_{n}=-{\e}2h(q^{2}_{n}+p^{2}_{n})\left(%
\begin{array}{c}
  hq_{n} \\
  q_{n}\\
  hp_{n} \\
  p_{n}\\
\end{array}
\right).
\]
The distance between stable and unstable manifolds can be
calculated by \bq
\partial_{\e}(W^{s}-W^{u})=M(\xi, h)u^{6}_{n},
\label{Mel}\eq
 where $M(\xi, h)$ is the Melnikov sum
\[
M(\xi, h)=\sum_{n=-\infty}^{\infty}\textbf{g}_{n}\cdot
\textbf{Z}_{n+1}
\]
After the substitution of homoclinic solutions, the Melnikov function
takes the form
\[
M(\xi,h)=-2h\sum_{n=-\infty}^{\infty}(Q^{2}_{n}+P^{2}_{n})\Big(Q_{n}\pt_{\xi}Q_{n}+P_{n}\pt_{\xi}P_{n}\Big).
\]
Let us consider
$$
{\tilde\alpha}_{1}=\frac{{\alpha}_{1}{\sinh}2W}{\sqrt{{\alpha}^{2}_{1}+{\alpha}^{2}_{2}}},\qquad
{\tilde\alpha}_{2}=\frac{{\alpha}_{2}{\sinh}2W
}{\sqrt{{\alpha}^{2}_{1}+{\alpha}^{2}_{2}}}.
$$
Then, the homoclinic solutions (\ref{solutions}) are given by
\[
Q_{n}(\xi)={\tilde\alpha}_{1}Q(\xi),\quad P_{n}(\xi)={\tilde
a}_{2}Q(\xi),\quad {\rm where}\quad Q(\xi)=\sech({\tilde W}n+\xi).
\]
It is clear that the Melnikov sum can be rewritten as
\[
M(\xi, h)=-\frac{h}{2}({\tilde\alpha}^{2}_{1}+
{\tilde\alpha}^{2}_{2})^{2}\pt_{\xi}\sum_{n=-\infty}^{\infty}Q^{4}.
\]
Using the Poisson summation formula we obtain
\begin{equation}\label{mel1}
M(\xi, h)=a_{\o}C_{_{M}}\sin ({2\pi\xi}/{\tilde W})+{\rm O}({\rm
e}^{-2{\pi}^{2}/{\tilde W}})
\end{equation}
where $$a_{\o}=({\tilde a}^{2}_{1}+{\tilde a}^{2}_{2})^{2},$$
$$C_{_{M}}(\tilde W)=4\frac{h}{2}{\pi}\big(\,\frac{2\pi}{\tilde W}\,\big)\big(\,
\frac{4}{3}\frac{\pi}{\tilde W}+\frac{4}{3}\frac{{\pi}^{3}}{{\tilde W}^{3}}\,\big){\rm e}^{-{\pi}^{2}/{\tilde W}}.$$

In order to calculate the Taylor series expansion of the Evans
function about the eigenvalue $\l=0$, we use the fact that $\l=0$
has algebraic multiplicity six. One must therefore calculate
${\pt}^{6}_{\l}E(0)$ (see Appendix):
%Furthermore, one has that
%the geometric multiplicity is four, and that each Jordan chain has
%length four.

\begin{equation}\label{Evans8}
{\pt}^{6}_{\l}E(0)=\frac{120\times 8}{D^{2}}{\rm B}_{1} {\rm
B}_{2} {\rm B}_{3}.
\end{equation}
where
\begin{equation}\label{Evans5}
{\rm B}_{1}=\sum_{n=-\infty}^{\infty}\frac{\alpha}{h}\,
\Big(\,Q^{c}_{n}{\pt}_{\xi}Q_{n}+P^{c}_{n}{\pt}_{\xi}P_{n}\,\Big)=\frac{1}{4}\int_{-\infty}^{\infty}
(Q^{2}(x)+P^{2}(x)){\id}x+{\rm O}(h),
\end{equation}
\begin{equation}\label{Evans6}
{\rm
B}_{2}=\sum_{n=-\infty}^{\infty} -\frac{\alpha}{h}\,Q_{n}{\pt}_{\o}Q_{n}=
-\frac{1}{2}{\pt}_{\o}\int^{\infty}_{\infty}Q^{2}(x){\id}x
\end{equation}
and
\begin{equation}\label{Evans7}
{\rm
B}_{3}=\sum_{n=-\infty}^{\infty} -\frac{\alpha}{h}\,P_{n}{\pt}_{\o}P_{n}=
-\frac{1}{2}{\pt}_{\o}\int^{\infty}_{\infty}P^{2}(x){\id}x+{\rm
O}(h).
\end{equation}
%Upon substituting the equations (\ref{Evans5}, \ref{Evans6},
%\ref{Evans7}) into the equation (\ref{Evans1}) one finally has
%that
%\begin{equation}\label{Evans8a}
%{\pt}^{6}_{\l}E(0)=\frac{120\times 8}{D^{2}}{\rm B}_{1} {\rm
%B}_{2} {\rm B}_{3}.
%\end{equation}

Due to the fact that CDNLS system is Hamiltonian, the eigenvalues
for the linearized problem will satisfy the relationship that if
$\l$ is an eigenvalue, then so is $-\l$ and $\pm{\l}^{*}$.
Furthermore, for the CDNLS $[0, 0, Q_{n}, P_{n}]^{\top}$ remains
an eigenfunction, and, in turn, $[{\pt}_{\o}Q_{n}, {\pt}_{\o}P_{n}, 0,
0]^{\top}$ remains a generalized eigenfunction; hence $\l=0$ is
an eigenvalue with multiplicity four in the perturbed problem.
Consequently, we need to
compute ${\pt}_{\e}{\pt}^{4}_{\l}E(0)$. One can write
\begin{equation}\label{Evans9}
{\pt}_{\e}{\pt}^{4}_{\l}E(0)=\bigg[\,{\pt}_{\e}(\textbf{Y}^{1,-}_{n}\wedge
\textbf{Y}^{1,+}_{n})\wedge
{\pt}^{2}_{\l}(\textbf{Y}^{2,-}_{n}\wedge
\textbf{Y}^{2,+}_{n})\wedge
{\pt}^{2}_{\l}(\textbf{Y}^{3,-}_{n}\wedge
\textbf{Y}^{3,+}_{n})\wedge \textbf{Y}^{1,+}_{n}\wedge
\textbf{Y}^{2,+}_{n}\wedge \textbf{Y}^{3,+}_{n}\,\bigg](0).
\end{equation}
As above and using $${\pt}_{\e}(\textbf{Y}^{1,-}_{n}\wedge
\textbf{Y}^{1,+}_{n})={\pt}_{\xi}M(\xi, h)u^{6}_{n},$$ we obtain
\begin{equation}\label{Evans10}
{\pt}_{\e}{\pt}^{4}_{\l}E(0)=\frac{4}{D^{2}}{\rm B}_{2}{\rm
B}_{3}{\pt}_{\xi}M(\xi, h).
\end{equation}
Combining the above results one has now an expansion of the Evans
function about $\l=0$ as
\begin{equation}\label{Evans11}
E(\l)={\rm B}_{2}{\rm
B}_{3}\frac{{\l}^{4}}{D^{2}}\Big[\, \frac{1}{6} {\e}{\pt}_{\xi}M(\xi,
h)+\frac{4}{3}{\l}^{2}{\rm B}_{1}\,\Big],
\end{equation}
where $M(\xi, h)$ is given by (\ref{mel1}).

The following theorem can then be stated.
%\newpage
\begin{thm}
Consider the stability of the solitary waves associated with the
CDNLS for $h>0$ sufficiently small. The associated linear operator
has four eigenvalues at $\l=0$. Furthermore, there exist only six
eigenvalues near $\l=0$. If $\xi=0$,
%{\tilde W}/2$,
then the wave is
linearly stable, and the two additional eigenvalues are purely
imaginary and are given to lowest order by
\[
{\l}^{\pm}_{s}={\pm}{\iu}\sqrt{\frac{\pi{\e}a_{\o}C_{_{M}}}{4
{\rm B}_{1} \tilde{W}}}
\]
whereas if $\xi={\tilde W}/2$, then the wave is linearly unstable, and the
two additional eigenvalues are real and are given to lowest order
by
\[
{\l}^{\pm}_{u}={\pm}\sqrt{\frac{\pi {\e}a_{\o}C_{_{M}}}{4
{\rm B}_{1} \tilde{W}}},
\]
%where ${\tilde\gamma}={2\pi}/{\tilde W}$.
\end{thm}

\begin{figure}[tbp]
\epsfxsize=8cm %\centerline{}
%\centerline{}
\epsffile{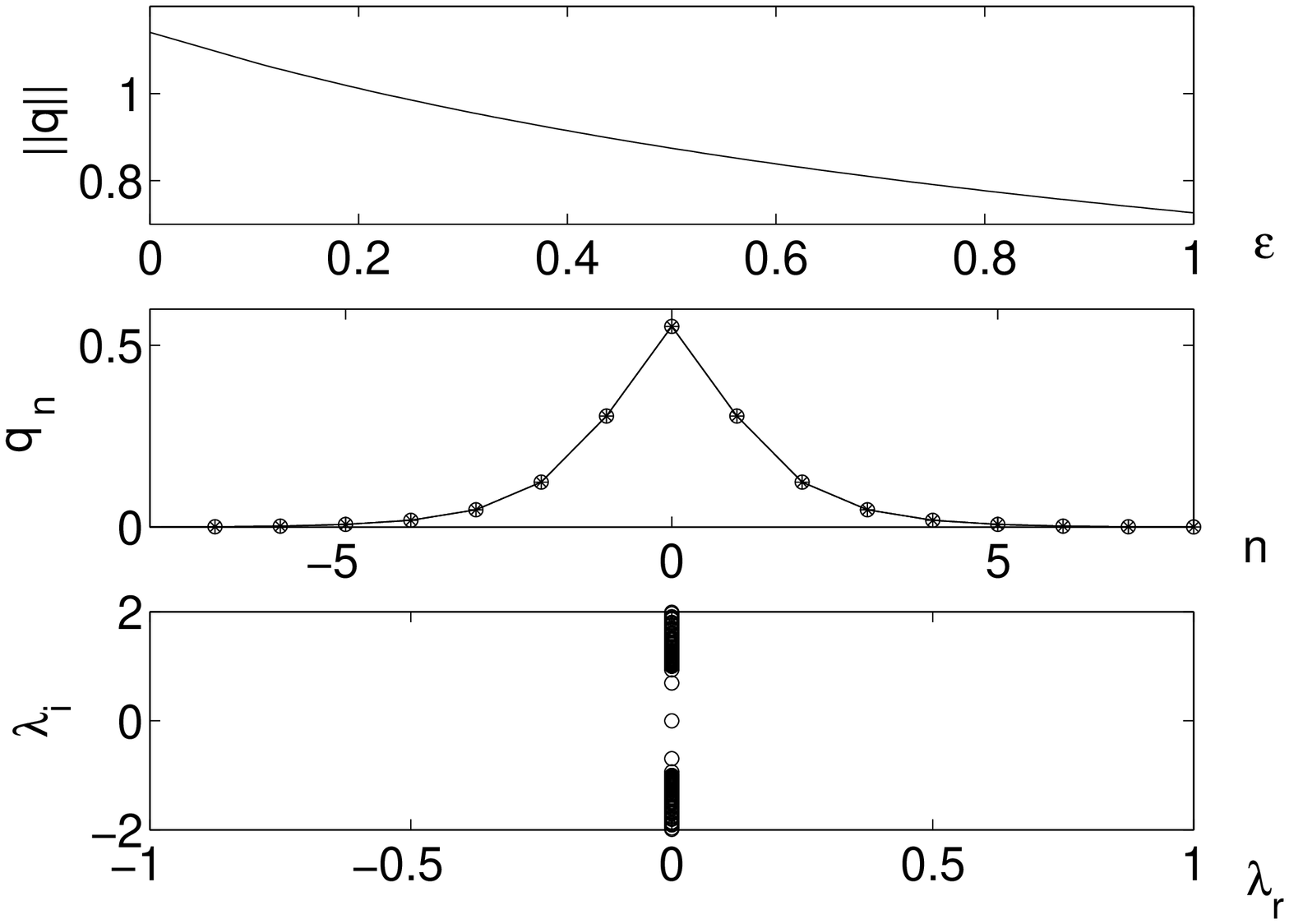}
\epsfxsize=8cm %\centerline{}
%\centerline{}
\epsffile{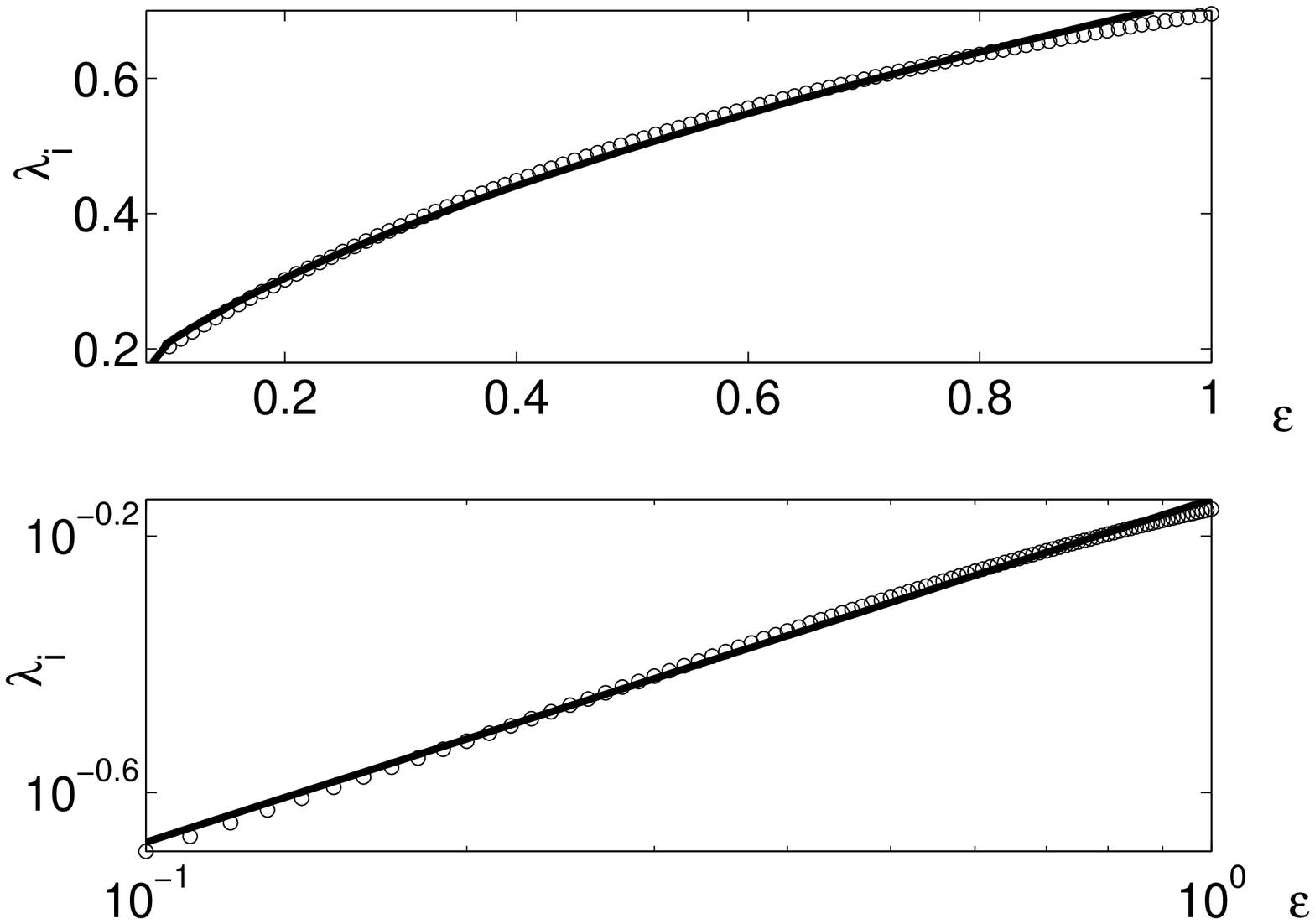}
\caption{The left panels of the figure show the branch of stable,
site-centered solitons stemming from the continuation of the integrable
case of $\epsilon=0$: the top panel shows the norm of the solution as
a function of $\epsilon$; the middle shows the discrete soliton spatial
profile for $\epsilon=1$ and the bottom panel shows the spectral plane
of the linearization eigenvalues around the solution (again for $\epsilon=1$).
The right panels show the evolution of the eigenvalue bifurcating from the
origin of the spectral plane (due to the breaking of the ``effective''
translational invariance of the integrable case. The trajectory of this
imaginary eigenvalue pair
is shown as a function of $\epsilon$ in linear (top panel)
and log-log (bottom panel) plot. The best fit power law is shown by solid
line and has the exponent $p=0.53$.}
\label{pfig1a}
\end{figure}

\begin{figure}[tbp]
\epsfxsize=8cm %\centerline{}
%\centerline{}
\epsffile{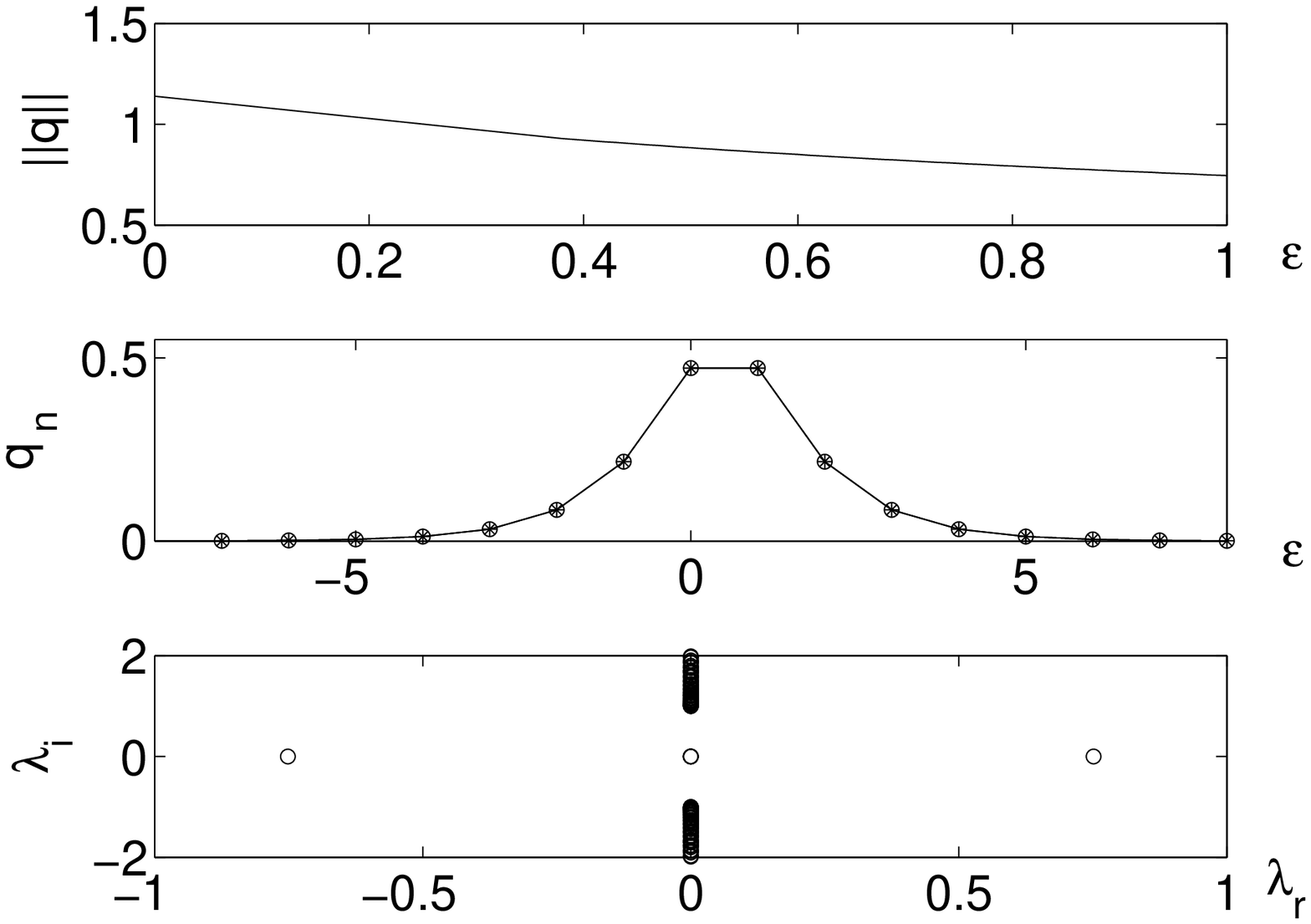}
\epsfxsize=8cm %\centerline{}
%\centerline{}
\epsffile{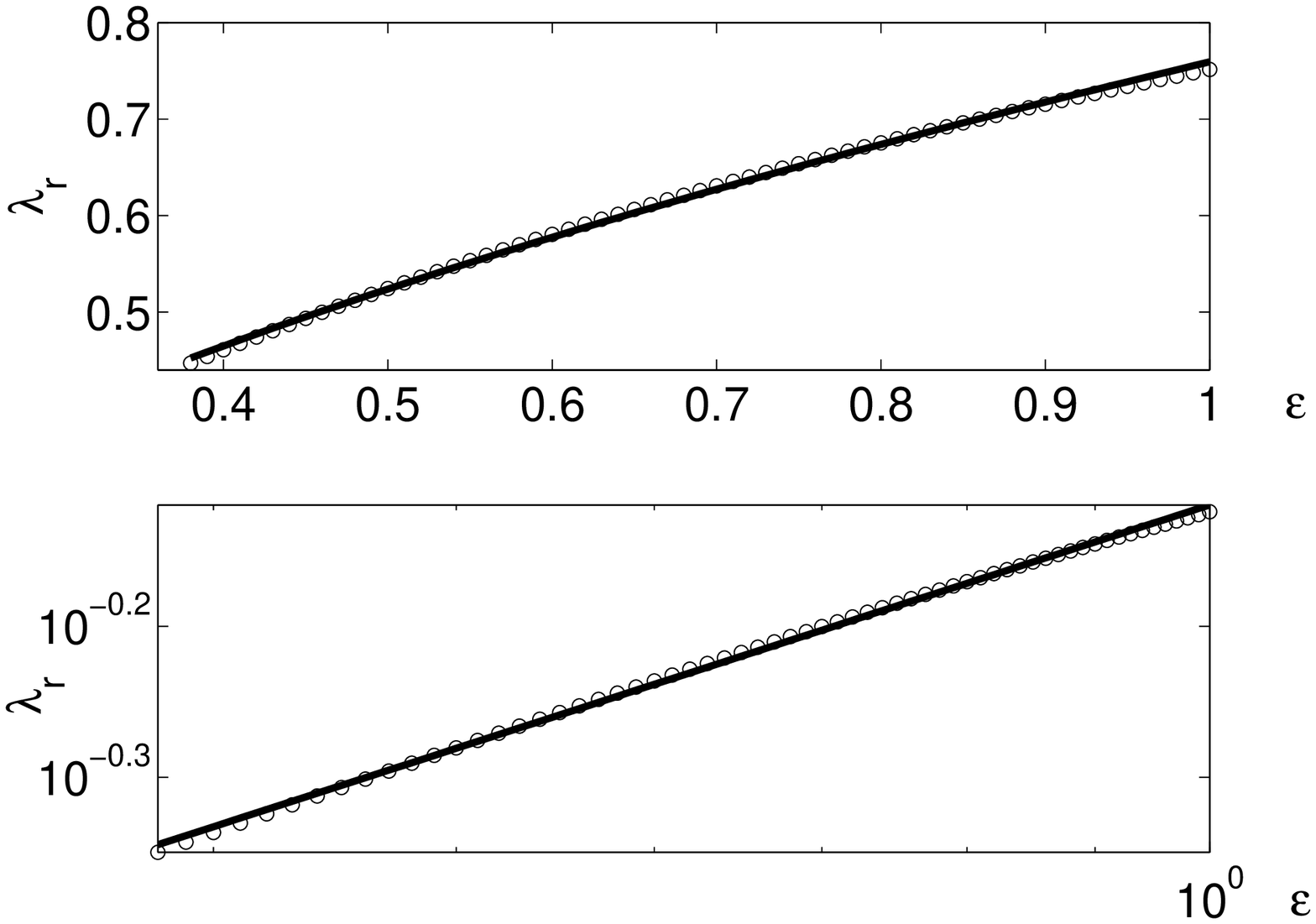}
\caption{The same features are shown as in Fig. 1 but for the inter-site
centered localized mode, where the relevant translational eigenvalue
pair becomes real rendering the configuration unstable.
The best fit exponent is $p=0.54$ in the right panel showing the real
eigenvalue pair as a function of $\epsilon$.}
\label{pfig2a}
\end{figure}

The predictions of Theorem 1 were numerically tested in Figs.
\ref{pfig1a} and \ref{pfig2a} for the site-centered and inter-site
centered modes numerically. The solutions were constructed for
different values of $\epsilon$, using a fixed point iteration
scheme on a $100$-site lattice.
The starting point used was the
analytically available solutions \cite{wadati} at the limit of
$\epsilon=0$ (i.e., the integrable limit).
Upon convergence to the exact numerical solution for
finite $\epsilon$, numerical linear
stability analysis was performed to obtain the spectrum of
linearization around the solitary waves.
We observed that
as soon as one deviates from the integrable limit, for this
nonlinearly coupled case, the effective translational invariance
of the discrete model is ``broken'' resulting in the bifurcation
of the relevant pair of eigenvalues from the origin of the
spectral plane. This bifurcation will occur along the imaginary
axis for the site-centered mode of Fig. \ref{pfig1a}, while
the eigenvalues will exit as real in the inter-site centered
case of Fig. \ref{pfig2a}, in accordance with Theorem 1.
Furthermore, the dependence of the eigenvalues will be
essentially proportional to $\sqrt{\epsilon}$ as is quantified
in the right panel of the relevant figures, which is consonant
with the prediction of Theorem 1 above, according to which
$\lambda \propto \epsilon^{1/2}$. On the contrary,
the two phase invariances of the two components, corresponding
to the respective ``mass'' conservation laws norm of each
field are preserved by the nonlinear coupling. As a result, the
relevant 2 pairs of eigenvalues at the spectral plane origin,
will remain at $\lambda=0$.

We note in passing that another implication of Theorem 1
concerns the near continuum limit behavior of the relevant
eigenvalues, according to which:
\begin{eqnarray}
\lambda \propto \exp \left(-\frac{\pi^2}{2 \sqrt{-\omega} h} \right),
\label{new}
\end{eqnarray}
for small $h$, which is consistent with the earlier findings
of \cite{kev5} (see also references therein about exponentially
small splittings of eigenvalues in the presence of discreteness).
Hence, approaching the continuum limit (for fixed $\epsilon$,
and $h \rightarrow 0$), the relevant translational eigenvalue pair would
approach $\lambda^2=0$, in direct analogy to its scalar case \cite{kev5}
(given this analogy, we do not examine this case further).
In the line of this qualitative analogy (the quantitative details and
relevant constants differ due to presence of two components here), 
Theorem 1 can be parallelized to Theorem 3.4 of \cite{kev5}.

A further note can be added regarding possible variations of the 
scattering lengths, which is relevant e.g., to our motivating
example of two hyperfine states of $^{87}Rb$ \cite{myatt,dsh}.
Slight deviations of the scatterings lengths from their unit 
values as well as possible changes of the coupling constants among
the different components do not change the relevant eigenvalue
count (since they do not affect the symmetries of the problem) or
the over-arching stability conclusions for on-site versus inter-site
modes. 

%%%%%%%%%%%%%%%%%%%%%%%%%%%%%%%%%%%%%%%

%%%%%%%%%%%%%%%%%%%%%%%%%%%%%%%%%%%%%%%

%%%%%%%%%%%%%%%%%%%%%%%%%%%%%%%%%%%%%%%

%%%%%%%%%%%%%%%%%%%%%%%%%%%%%%%%%%%%%%%

\section{Coupled DNLS with Linear Coupling}

In this section, we consider a system of CDNLS equations with linear
coupling:
\begin{eqnarray}
{\iu}{\dot
  u}_{n}&=&-{\D}_{2}u_{n}-(\abs{u_{n}}^{2}+\abs{v_{n}}^{2})(u_{n+1}+u_{n-1}) +
{\delta}v_{n}\cr\cr {\iu}{\dot
  v}_{n}&=&-{\D}_{2}v_{n}-(\abs{v_{n}}^{2}+\abs{u_{n}}^{2})(v_{n+1}+v_{n-1}) +
{\delta}u_{n}. \label{lcdnls}
\end{eqnarray}

We are interested in stationary solutions of the system of
equations (\ref{cdnls}) and make the ansatz

\bq u_{n}(t)=q_{n}{\rm e}^{-{\iu}{\o}t}, \qquad v_{n}(t)=p_{n}{\rm
e}^{-{\iu}{\o}t}, \label{stationary1} \eq

with a uniform rotation frequency $\o$. Upon setting
\[
r_{n}=(q_{n}-q_{n-1})/h,\quad s_{n}=(p_{n}-p_{n-1})/h,
\]
the steady--state problem for CDNLS can be formulated as

\bqn q_{n+1}&=&({\rm
c}{\chi}-1)q_{n}+r_{n}h+ {\delta} h^2 p_{n} {\chi}\cr\cr
r_{n+1}&=&({\rm
c}{\chi}-2)\frac{q_{n}}{h}+r_{n}+ \delta {h} p_{n} {\chi}\cr\cr
p_{n+1}&=&({\rm
c}{\chi}-1)p_{n}+s_{n}h+{\delta} h^2 q_{n}{\chi}\cr\cr
s_{n+1}&=&({\rm
c}{\chi}-2)\frac{p_{n}}{h}+s_{n}+{\delta} {h} q_n {\chi}
\label{map21} \eqn where
\[
{\rm c}=2-{\o}h^{2}\quad{\rm and}\quad
{\chi}=\frac{1}{1+h^{2}(q^{2}_{n}+p^{2}_{n})}.
\]

The equations (\ref{map21}) are written so that, when $\delta=0$,
they revert to the steady--state problem for integrable CDNLS.

The equation of variation with respect to $\e$ for the stable
$W^{s}$ and $W^{u}$ manifolds is given (as in the CDNLS
(\ref{cdnls}) by the non-homogeneous problem \bq \textbf{Y}_{n+1}=
\textbf{A}(\l, n)\textbf{Y}_{n}+\textbf{g}_{n}, \label{nonhom1} \eq
where $\{\,\textbf{g}_{n}\,\}$ is a uniformly bounded sequence:
\[
\textbf{g}_{n}=\delta h \left(%
\begin{array}{c}
  h p_{n} \chi \\
  p_{n} \chi \\
  h q_{n} \chi\\
  q_{n} \chi.\\
\end{array}
\right)
\]

The distance between stable and unstable manifolds can be
calculated by \bq
\partial_{\delta}(W^{s}-W^{u})=M_2(\omega, h)u^{4}_{n}.
\label{lMel}\eq
$M_{2}(\omega, h)$ is the Melnikov sum
\[
M_{2}(\omega, h)=\sum_{n=-\infty}^{\infty}\textbf{g}_{n}\cdot {\tilde
\textbf{Z}}_{n+1},
\]
where the factor $\chi$ in $\textbf{g}$ is eliminated because we
examine what happens for small $h$ and
$${\tilde \textbf{Z}}_{n}=\Big[\,\pt_{\omega}(Q_{n-1}-Q_{n})/h,
\pt_{\omega}Q_{n}, \pt_{\omega}(P_{n-1}-P_{n})/h,
\pt_{\omega}P_{n}\,\Big]^{\top}$$ the second ``growth mode" is
obtained by solving the inhomogeneous equation.

It is clear that the Melnikov sum can be rewritten as
\[
M_{2}(\omega,
h)={h}{\delta}\pt_{\omega}\sum_{n=-\infty}^{\infty}P_{n}Q_{n}
=\delta \partial_{\omega} \int_{-\infty}^{\infty} P(x) Q(x) dx +{\rm
O}(h),
\]
where $Q_{n}, P_{n}$ are defined in (\ref{solutions}).
%Using the
%Poisson summation formula we obtain
%\begin{equation}\label{mel2} M_{2}(\omega, h)=h{\delta}{\pt}_{\omega}\Big[\, A(\omega)+A(\omega)(2h/{\tilde W})
%\sum_{n}C_{n}(\omega){\cos}\frac{2\pi\xi}{\tilde W}\,\Big]
%\end{equation}
%where $${\tilde
%W}={\cosh}^{-1}\big(\,1-\frac{{\omega}h^{2}}{2}\,\big),\quad
%A(\omega)=(2h/{\tilde W}) {\sinh}^{2}2{\tilde W}$$
%$$
% C_{n}(\omega)=2{\pi}\big(\,\frac{2{\pi}n}{\tilde W}\,\big){\rm csch}\frac{{\pi}^{2}n}{\tilde W}$$
Using the continuum limit of $P$ and $Q$, one can approximate
$M_2$ close to the continuum limit as:
\begin{eqnarray}
M_2=-\frac{\alpha_1 \alpha_2}{\alpha_1^2+\alpha_2^2} \delta
    (-\omega)^{-1/2}.
\label{add1}
\end{eqnarray}

Combining the above results one has now an expansion of Evans
function about $\l=0$, similarly to Eq. (\ref{Evans11}),
involving the derivatives ${\pt}_{\delta}{\pt}^{4}_{\l} (0)$
and ${\pt}^{6}_{\l} E(0)$, as the leading terms; the details
are left to the interested reader, being quite similar to those
of the previous section. One then has the following theorem:

%\begin{equation}\label{EvansL}
%E_{2}(\l)={\rm B}_{2}{\rm
%B}_{3}\frac{{\l}^{4}}{D^{2}}\Big[\,{\delta}{\pt}_{\omega}M_{2}(\omega,
%h)+\frac{4}{3}{\l}^{2}{\rm B}_{1}\,\Big]
%\end{equation}
%where $M_{2}(\xi, h)$ is given by (\ref{add1}) and ${\rm B}_{1},
%{\rm B}_{2}, {\rm B}_{3}, {\rm B}_{4}$ are given by (\ref{Evans5},
%\ref{Evans6}, \ref{Evans7}). For $h\ll 1$, we have
%$\tilde{W}=\sqrt{-\omega} h$, then $B_1 \approx \alpha_2^2
%(-\omega)^{(-1/2)} /(2 (\alpha_1^2+\alpha_2^2))$.
%%the first and second term in
%%the bracket of (\ref{EvansL}) are negative, for every negative
%%value of $\omega$.
%\textbf{in (4.7) inside the bracket $B_3$ has been replaced by
%$B_1$}
%We state the following theorem:

\begin{thm}
Consider the stability of the solitary waves associated with the
CDNLS (\ref{lcdnls}) for $h>0$ sufficiently small. The associated
linear operator has four eigenvalues at $\l=0$. Furthermore, there
exists an additional pair of eigenvalues near $\l=0$ which
are purely imaginary and directly proportional to $\delta$.
\end{thm}

%%%%%%%%%%%%%%%%%%%%%%%%%%%%%%%%%%%%%%%

\begin{figure}[tbp]
\epsfxsize=8cm %\centerline{}
%\centerline{}
\epsffile{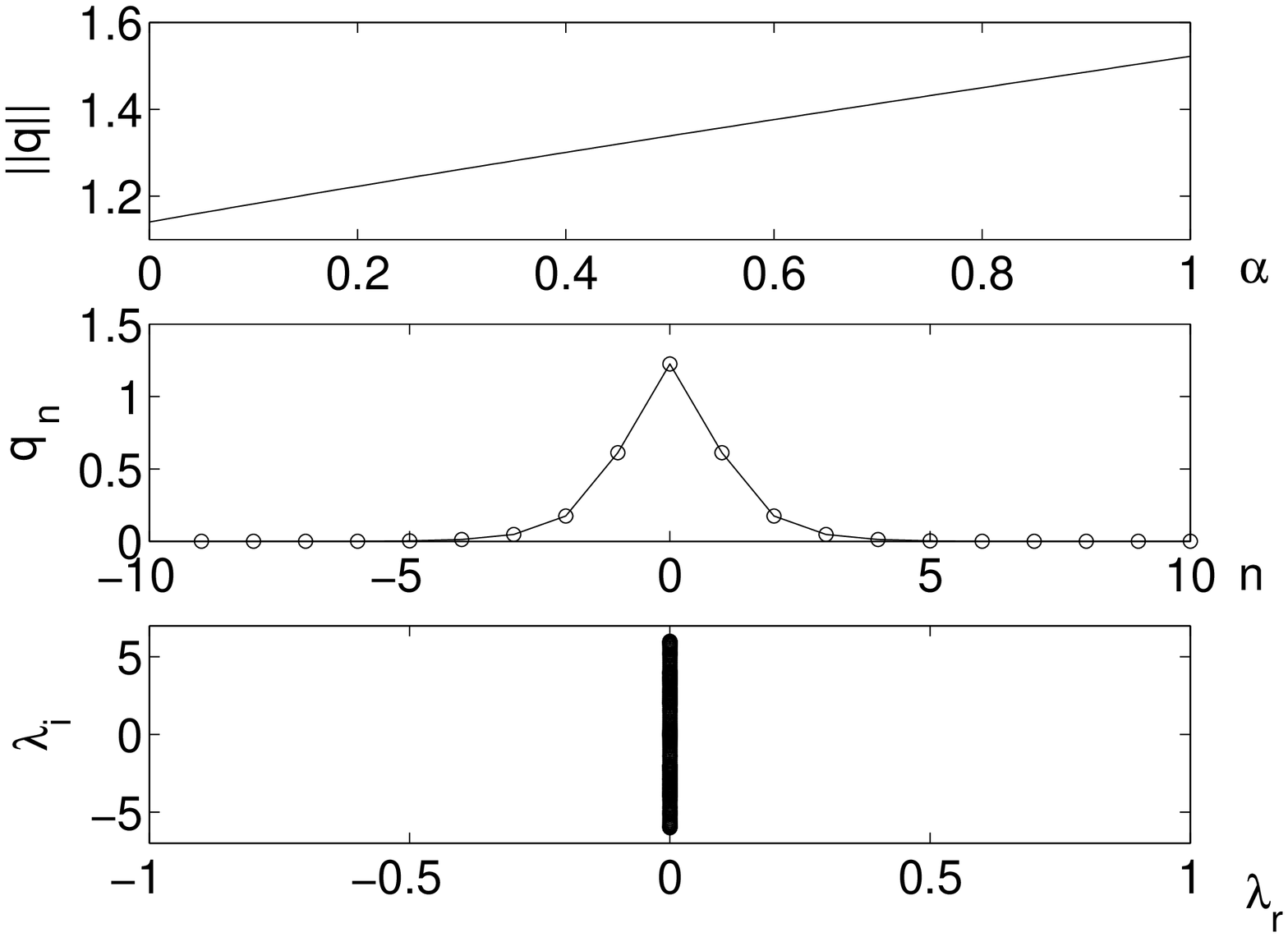}
\epsfxsize=8cm %\centerline{}
%\centerline{}
\epsffile{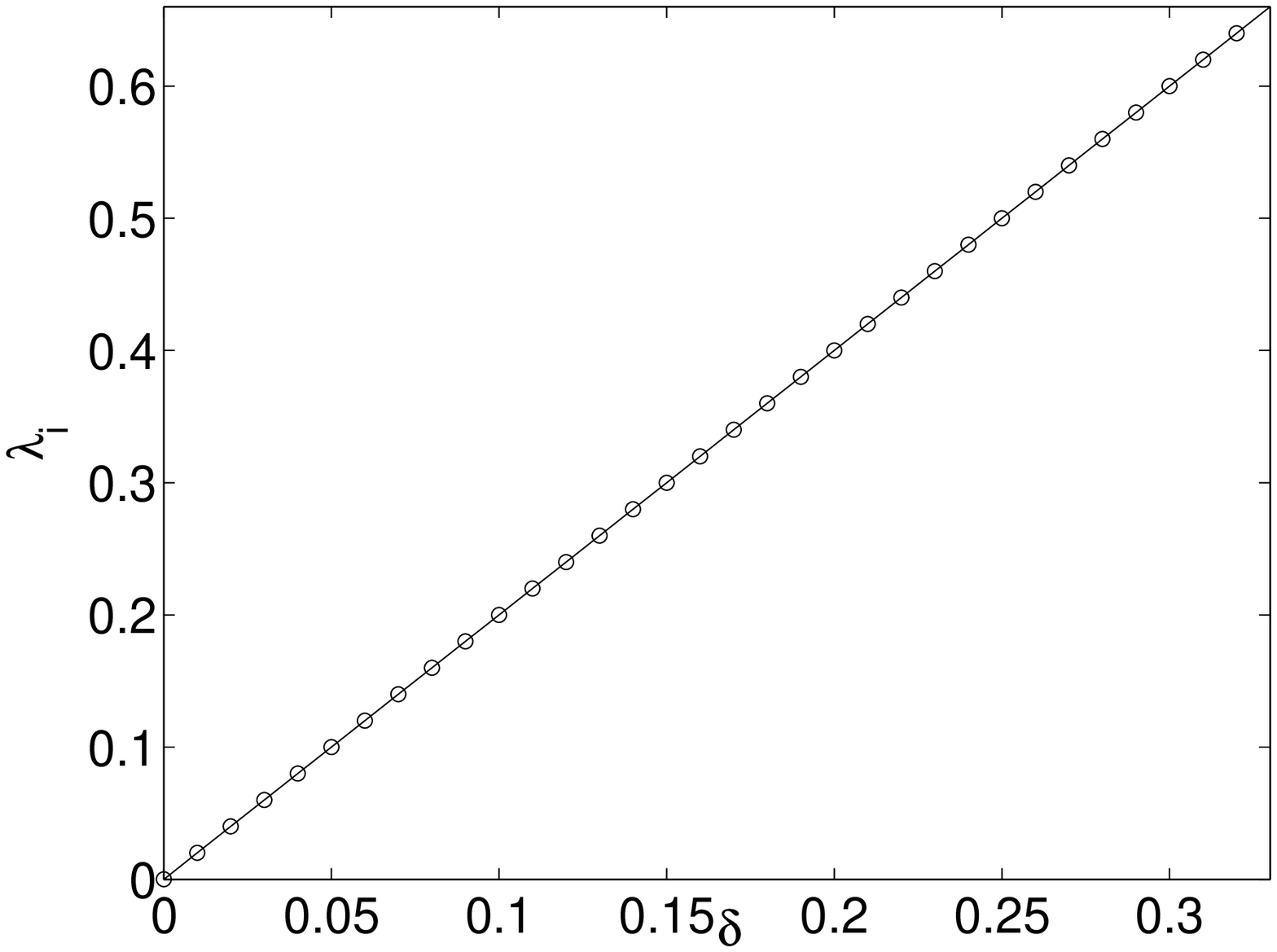}
\caption{The left panels are similar to those of Fig. \ref{pfig1a},
but are now for the site-centered mode in the case of linear coupling
between the components. The middle and bottom left panels show the
field profile and the stability for the case of $\delta=1$. The right
panel shows the eigenvalue bifurcating from the origin as soon as
$\delta$ becomes non-zero. The circles denote the full numerical results,
while the solid line denotes the curve $\lambda_i=2 \delta$, which
approximates very well the numerical result. }
\label{pfig3a}
\end{figure}

%%%%%%%%%%%%%%%%%%%%%%%%%%%%%%%%%%%%%%%

%%%%%%%%%%%%%%%%%%%%%%%%%%%%%%%%%%%%%%%

\begin{figure}[tbp]
\epsfxsize=8cm %\centerline{}
%\centerline{}
\epsffile{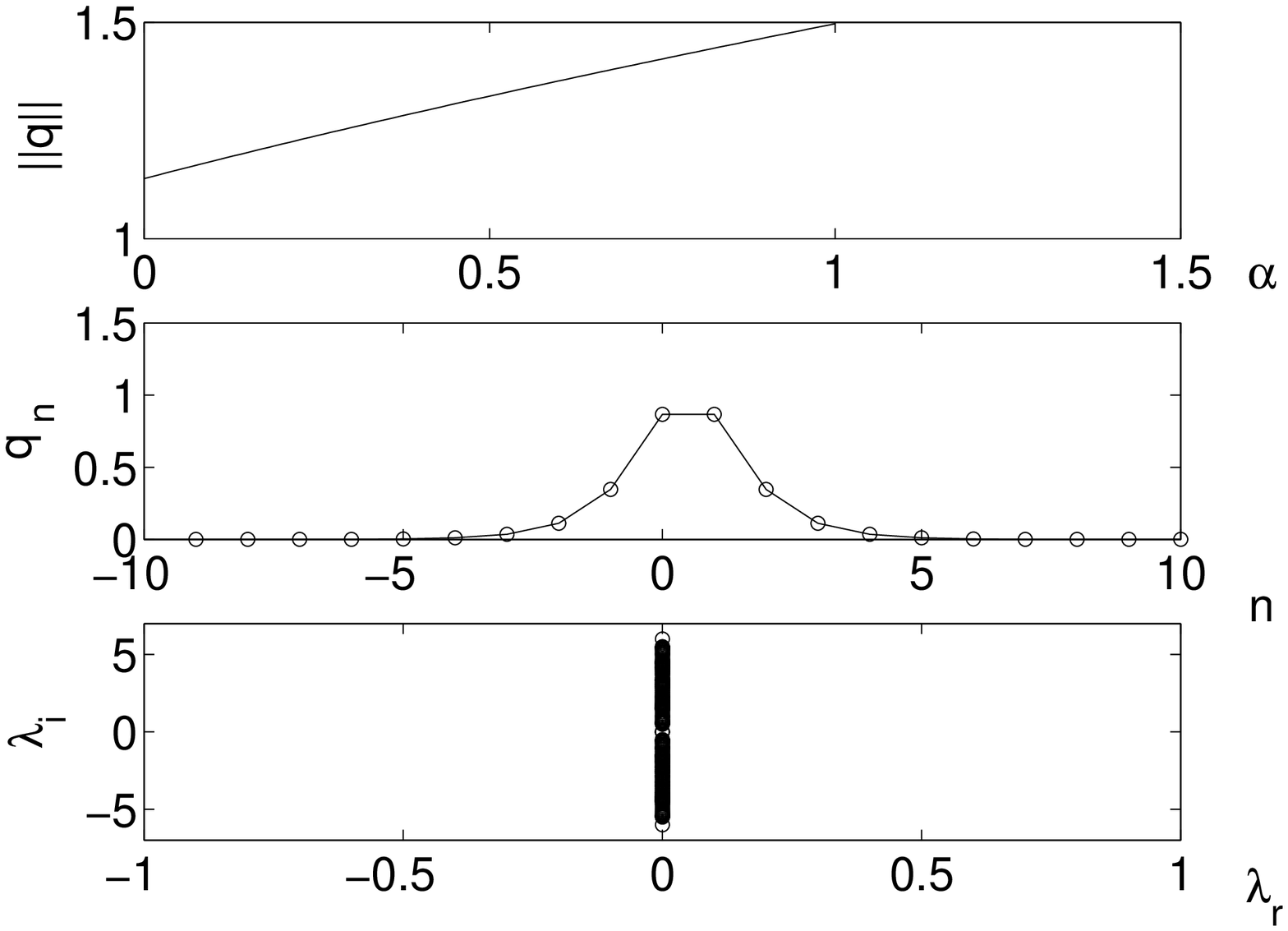}
\epsfxsize=8cm %\centerline{}
%\centerline{}
\epsffile{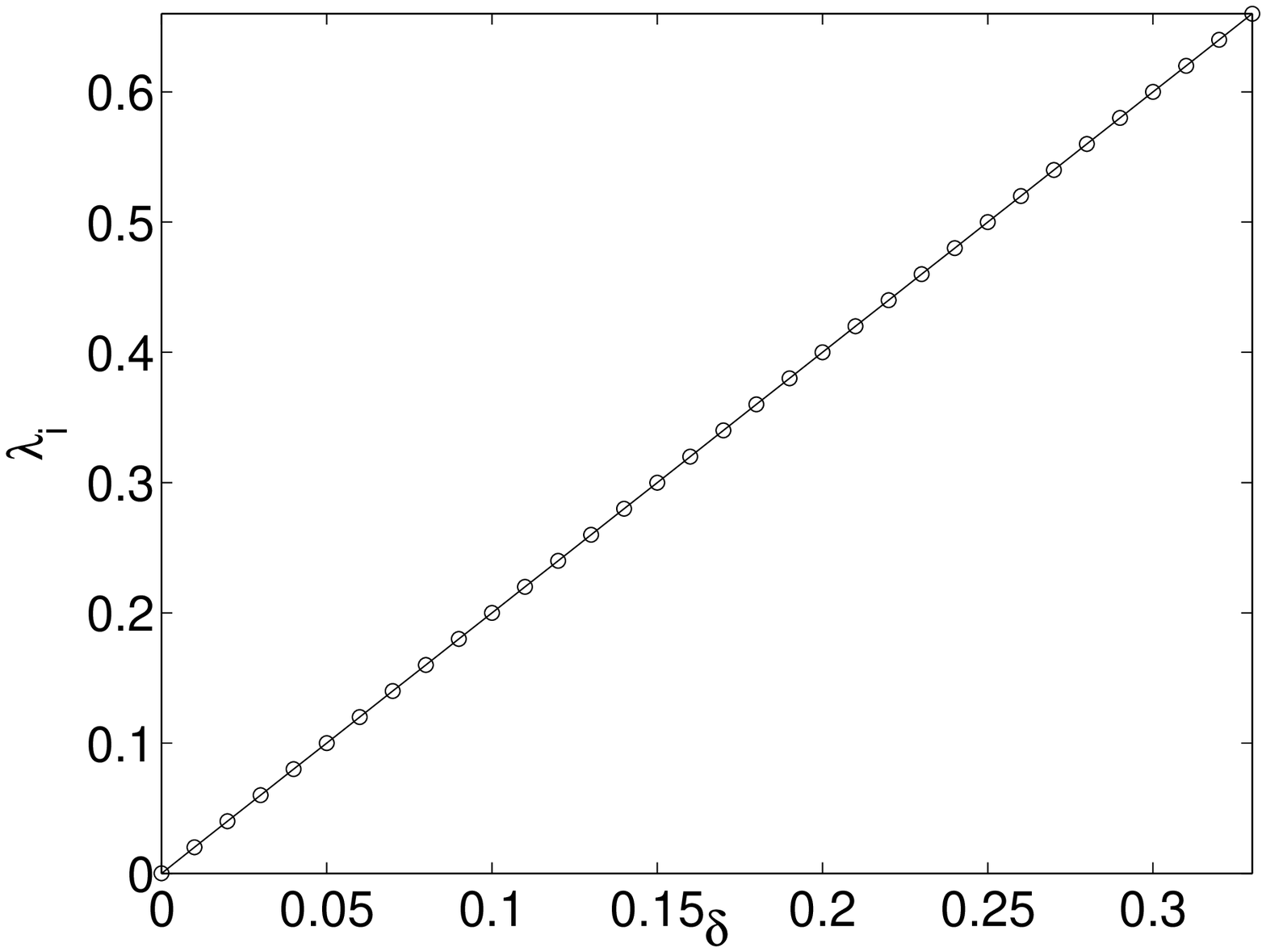}
\caption{The same features are shown as in Fig. \ref{pfig3a} but
for the case of the inter-site centered mode. The middle and bottom left panels
are for $\delta=0.5$. The right panel shows the eigenvalue bifurcating
from the origin. The circles denote the full numerical linear stability
results, while the solid line denotes the curve $\lambda_i=2 \delta$.}
\label{pfig4a}
\end{figure}

%%%%%%%%%%%%%%%%%%%%%%%%%%%%%%%%%%%%%%%

%\subsection{Remaining Issue-old versions}

We have tested the above predictions numerically
and have found them to be in good agreement with the
computational results. In particular, the site-centered and
inter-site centered modes are respectively shown in Figs. \ref{pfig3a}
and \ref{pfig4a} for the case of linear coupling (in order
to compare/contrast them with those of nonlinear coupling in
Figs. \ref{pfig1a}-\ref{pfig2a}). The main thing to notice in this
case is that the eigenvalue bifurcation is not only linear
in its dependence of the linear coupling parameter $\alpha$,
but is furthermore along the imaginary axis for  {\it both}
the site-centered and inter-site centered solutions. The latter
implies the absence of instability in such linearly coupled cases.
It is important to highlight here a key difference between the
linear and nonlinear coupling case, to which this difference in
stability properties can be attributed. In the latter case,
examined previously, the effective translational invariance of
the integrable limit was ``destroyed'' upon the action of the nonlinear
coupling, while the relevant phase invariances remained intact.
On the contrary, in the linearly coupled case, the translational
invariance remains present, while one of the phase invariances
is destroyed, as it is now only the sum of the squared $l^2$ norms
that is conserved (rather than each individual one of them).
As a result, the source of the bifurcation, and hence the ensuing
stability behavior is different in the two cases. This is also implicitly
mirrored in the similar stability behavior of the site-centered
and inter-site centered modes. Finally, let us make, in passing,
another important observation for the linear case. For the setting
examined above with solutions identical between the two components
i.e., with $q_n=p_n$, it is straightforward to note that the
exact solution is analytically available for all values of the
linear coupling $\delta$, as the latter merely renormalizes
$W$, by replacing $\omega$ with $\omega-\delta$ in the relevant
expression. Furthermore, an alternative way to see that the case
with the linear coupling should lead to linear dependence of the
near $\lambda=0$ eigenvalue pair as a function of $\delta$ is
the following. Consider the linear transformation \cite{bernard}
\[
\left(%
\begin{array}{c}
  u_n \\
  v_n \\
\end{array}
\right)
=\left(
\begin{array}{cc}
\cos(\delta t)& -i \sin(\delta t)\\
-i \sin(\delta t) & \cos(\delta t)
\end{array}
\right)
\left(%
\begin{array}{c}
  \tilde{u}_n \\
  \tilde{v}_n \\
\end{array}
\right).
\]
It is interesting to observe that the equations satisfied by
$\tilde{u}_n$ and $\tilde{v}_n$ are those of the {\it original}
model i.e., {\it without the linear coupling}. Hence, the
linear coupling can be ``factored out'' through this transformation,
which is motivated by first-order
vector systems of ordinary differential equations (and respected
by our Manakov-type nonlinearity). The eigenvalues of the
transformation matrix are $\exp(i \delta t)$ and  $\exp(-i \delta t)$,
which in turn suggests a linear dependence of
the bifurcating pair of eigenvalues on $\delta$ in agreement with
our numerical results of Figs. \ref{pfig3a}-\ref{pfig4a}.

%=(1/2) {\rm acosh}(1-(\omega-\alpha)/2)$ (for $h=C=1$)

\section{Conclusions}

In conclusion, in this paper we have developed the Evans function
methodology for discrete nonlinear Schr{\"o}dinger equations in the
vector case. We have used as our starting point the integrable
discretization of \cite{ablowitz} and have introduced nonlinear,
as well as linear perturbations to it breaking different kinds of
invariances. As a result, pairs of eigenvalues, corresponding
respectively to
these invariances (of the linearization around solitary wave solutions of the
equation) have moved away from the origin of the spectral
plane. These pairs have been analytically tracked via the
zeros of the Evans function and have been found to yield critical
information about the stability of the solutions in the non-integrable,
perturbed case.

In the nonlinearly coupled case, an eigenvalue pair
corresponding to translations
of the solitary waves has been found to bifurcate from 0, leading to
either a stable (site-centered) or unstable (inter-site centered) solution.
On the other hand, in the linearly coupled case, a pair corresponding
to the phase invariance of the waves has been found to bifurcate linearly
from the origin. In both settings, the computed results
based on numerical existence and linear stability methods have successfully
corroborated the analytical predictions.

It would be interesting to extend the methodology presented herein
to the, very intensely studied in recent years, context of
photorefractive materials \cite{moti,review}. In the latter case,
the nonlinearity is of the saturable type (coinciding with the
ones studied above for low intensities, but having very different
-quasi-linear- behavior at high intensities). Such studies and the
corresponding extensions to higher-dimensional settings are
currently in progress and will be reported in future publications.

\vspace{5mm}

{\bf Acknowledgements.} PGK gratefully acknowledges support from
NSF-DMS-0204585, NSF-DMS-0505663, NSF-DMS-0619492
and NSF-CAREER. VMR acknowledges support from IACM-FORTH, Greece.

\appendix

\section{Calculation of ${\pt}^{6}_{\l}E(0)$. }

Here, we calculate the ${\pt}^{6}_{\l}E(0)$. Following Kapitula
\cite{kapitula99}, \cite{kev5} one has
\begin{equation}\label{Evans1}
{\pt}^{6}_{\l}E(0)=120\bigg[\,{\pt}^{2}_{\l}(\textbf{Y}^{1,-}_{n}\wedge
\textbf{Y}^{1,+}_{n})\wedge
{\pt}^{2}_{\l}(\textbf{Y}^{2,-}_{n}\wedge
\textbf{Y}^{2,+}_{n})\wedge
{\pt}^{2}_{\l}(\textbf{Y}^{3,-}_{n}\wedge
\textbf{Y}^{3,+}_{n})\wedge \textbf{Y}^{1,+}_{n}\wedge
\textbf{Y}^{2,+}_{n}\wedge \textbf{Y}^{3,+}_{n}\,\bigg](0).
\end{equation}
For $i=1, 2, 3$ we have
\begin{equation}\label{Evans2}
{\pt}^{2}_{\l}\textbf{Y}^{i,\pm}_{n+1}(0)=\textbf{A}(0,
n){\pt}^{2}_{\l}\textbf{Y}^{i,\pm}_{n}(0)+2{\pt}_{\l}\textbf{A}(0,
n){\pt}_{\l}\textbf{Y}^{i,\pm}_{n}(0),
\end{equation}
where
\begin{eqnarray}
\label{Evans3} {\pt}_{\l}\textbf{Y}^{1,\pm}_{n}(0)&=&\Big[\,0,
Q^{c}_{n}, 0, (Q^{c}_{n}-Q^{c}_{n-1})/h, 0, P^{c}_{n}, 0,
(P^{c}_{n}-P^{c}_{n-1})/h\,\Big]^{\top}\cr\cr
{\pt}_{\l}\textbf{Y}^{2,\pm}_{n}(0)&=&\Big[\,{\pt}_{\o}Q_{n}, 0,
{\pt}_{\o}(Q_{n}-Q_{n-1})/h, 0, 0, 0, 0, 0\, \Big]^{\top}\cr\cr
{\pt}_{\l}\textbf{Y}^{3,\pm}_{n}(0)&=&\Big[\,0, 0, 0, 0,
{\pt}_{\o}P_{n}, 0, {\pt}_{\o}(P_{n}-P_{n-1})/h, 0, \,\Big]^{\top}
\end{eqnarray}
Set $\textbf{g}^{i}_{n}={\pt}_{\l}\textbf{A}(0,
n){\pt}_{\l}\textbf{Y}^{i,\pm}_{n}(0)$ for $i=1, 2, 3$. Upon using
the variation of parameters to solve equation (\ref{Evans2}), one
has that
\begin{eqnarray}
\label{Evans3a} {\pt}^{2}_{\l}(\textbf{Y}^{1,-}_{n}\wedge
\textbf{Y}^{1,+}_{n})(0)&=&\frac{2}{D}\bigg(\,
u^{6}_{n}\sum_{n=-\infty}^{\infty}\textbf{g}^{1}_{n}\cdot
\textbf{Z}^{1}_{n+1}+c_{1}u^{1}_{n}+c_{2}u^{2}_{n}+c_{3}u^{3}_{n}\,\bigg)\cr\cr
{\pt}^{2}_{\l}(\textbf{Y}^{2,-}_{n}\wedge
\textbf{Y}^{2,+}_{n})(0)&=&\frac{2}{D}\bigg(\,
u^{5}_{n}\sum_{n=-\infty}^{\infty}\textbf{g}^{2}_{n}\cdot
\textbf{Z}^{2}_{n+1}+c_{4}u^{1}_{n}+c_{5}u^{2}_{n}+c_{6}u^{3}_{n}\,\bigg)\cr\cr
{\pt}^{2}_{\l}(\textbf{Y}^{3,-}_{n}\wedge
\textbf{Y}^{3,+}_{n})(0)&=&\frac{2}{D}\bigg(\,
u^{4}_{n}\sum_{n=-\infty}^{\infty}\textbf{g}^{3}_{n}\cdot
\textbf{Z}^{3}_{n+1}+c_{7}u^{1}_{n}+c_{8}u^{2}_{n}+c_{9}u^{3}_{n}\,\bigg).
\end{eqnarray}
Here $c_{1},\cdots, c_{9}$ are constants. The above equations can
be simplified as follows

\begin{eqnarray}
\label{Evans4} {\pt}^{2}_{\l}(\textbf{Y}^{1,-}_{n}\wedge
\textbf{Y}^{1,+}_{n})(0)&=&\frac{2}{D}\bigg(\,
u^{6}_{n}\sum_{n=-\infty}^{\infty}\frac{\alpha}{h}\,
\Big(\,Q^{c}_{n}{\pt}_{\xi}Q_{n}+P^{c}_{n}{\pt}_{\xi}P_{n}\,\Big)+
c_{1}u^{1}_{n}+c_{2}u^{2}_{n}+c_{3}u^{3}_{n}\,\bigg)\cr\cr
{\pt}^{2}_{\l}(\textbf{Y}^{2,-}_{n}\wedge
\textbf{Y}^{2,+}_{n})(0)&=&\frac{2}{D}\bigg(\,
u^{5}_{n}\sum_{n=-\infty}^{\infty}
-\frac{\alpha}{h}\,Q_{n}{\pt}_{\o}Q_{n}
+c_{4}u^{1}_{n}+c_{5}u^{2}_{n}+c_{6}u^{3}_{n}\,\bigg)\cr\cr
{\pt}^{2}_{\l}(\textbf{Y}^{3,-}_{n}\wedge
\textbf{Y}^{3,+}_{n})(0)&=&\frac{2}{D}\bigg(\,
u^{4}_{n}\sum_{n=-\infty}^{\infty} -
\frac{\alpha}{h}\,P_{n}{\pt}_{\o}P_{n}
+c_{7}u^{1}_{n}+c_{8}u^{2}_{n}+c_{9}u^{3}_{n}\,\bigg).
\end{eqnarray}

\end{document}